\documentclass[final]{IEEEtran}

\usepackage{amsthm,amssymb,graphicx,multirow,amsmath,color,amsfonts}
\usepackage[update,prepend]{epstopdf}
\usepackage[noadjust]{cite}
\usepackage[latin1]{inputenc}
\usepackage{tikz}
\usetikzlibrary{arrows,calc} 
\usepackage{bbm} 
\usepackage{pdfpages}
\usepackage{tabulary}
\usepackage{multirow}
\usepackage{comment}
\usepackage{setspace}
\usepackage{subfigure}

\def\nbo{{\mathbf{o}}}

\def\nbx{{\mathbf{x}}}
\def\nby{{\mathbf{y}}}

\def\nb0{{\mathbf{0}}}
\def\nb1{{\mathbf{1}}}

\def\ncalC{{\mathcal{C}}}

\def\ncalR{{\mathcal{R}}}
\def\ncalS{{\mathcal{S}}}

\def\ncalX{{\mathcal{X}}}

\def\nbbN{{\mathbb{N}}}

\def\nbbP{{\mathbb{P}}}

\newtheorem{lemma}{Lemma}

\newtheorem{theorem}{Theorem}

\newtheorem{cor}{Corollary}

\newtheorem{remark}{Remark}

\def\argmin{\operatorname{arg~min}}

\allowdisplaybreaks

\begin{document}
\graphicspath{{./Figures/}}
\title{Handover Probability in Drone Cellular Networks
}
\author{
Morteza Banagar, \textit{Student Member, IEEE}, Vishnu Vardhan Chetlur, \textit{Student Member, IEEE}, and \\Harpreet S. Dhillon, \textit{Senior Member, IEEE}
\thanks{The authors are with Wireless@VT, Department of ECE, Virginia Tech, Blacksburg, VA (email: \{mbanagar, vishnucr, hdhillon\}@vt.edu). The support of the US NSF (Grants CNS-1617896 and CNS-1923807) is gratefully acknowledged.}
}

\maketitle

\vspace{-1.7cm}
\begin{abstract}
This letter analyzes the handover probability in a drone cellular network where the initial positions of drone base stations (DBSs) serving a set of user equipment (UE) on the ground are modeled by a homogeneous Poisson point process (PPP). Inspired by the mobility model considered in the third generation partnership project (3GPP) studies, we assume that all the DBSs move along straight lines in random directions. We further consider two different scenarios for the DBS speeds: (i) same speed model (SSM), and (ii) different speed model (DSM). Assuming nearest-neighbor association policy for the UEs on the ground, we characterize the handover probability of this network for both mobility scenarios. For the SSM, we compute the exact handover probability by establishing equivalence with a single-tier terrestrial cellular network, in which the base stations (BSs) are static while the UEs are mobile. We then derive a lower bound for the handover probability in the DSM by characterizing the evolution of the spatial distribution of the DBSs over time.
\end{abstract}

\vspace{-0.2cm}
\section{Introduction} \label{sec:intro}
The use of drone-mounted BSs for providing wireless connectivity has emerged as a promising solution to supplement the coverage and capacity of terrestrial cellular networks. While the mobility of the DBSs offers numerous benefits in many scenarios, it also drastically changes the cellular architecture from the one with carefully placed and reliable terrestrial BSs to the one with mobile and often short-lived DBSs \cite{J_Mozaffari_Tutorial_2018}. One of the key consequences of the mobility of DBSs is the occurrence of handovers, even if the UEs are static on the ground. Since handovers result in signaling overhead, it is highly desirable to carefully understand the handover behavior in this new operational regime. In addition, since handovers have traditionally been studied by assuming mobile UEs and static BSs, it is also natural to wonder whether there is some underlying connection between the statistics of the handovers observed in these two fundamentally different regimes. Inspired by such questions, we present a rigorous analysis of the handover probability in drone cellular networks using tools from stochastic geometry \cite{B_Haenggi_Stochastic_2012}.

{\em Prior Art.} The handover probability is a well-investigated metric in terrestrial cellular networks with static BSs and mobile UEs \cite{J_Sadr_Handoff_2015, J_Shankar_Spatio_2017, J_Bao_Stochastic_2015, J_Arshad_Velocity_2017}. Modeling the locations of BSs as a 2D homogeneous PPP, the authors in \cite{J_Sadr_Handoff_2015} derived the probability of the occurrence of the first handover for a reference UE that moves at a constant speed along a straight line. The authors of \cite{J_Shankar_Spatio_2017} have also derived this result in the analysis of the joint coverage probability of cellular networks. In \cite{J_Bao_Stochastic_2015}, the authors analyzed the horizontal and vertical handoff rates in multi-tier heterogeneous networks for arbitrary trajectories of UEs. While these works are useful for designing and optimizing terrestrial networks, they are not directly applicable to drone cellular networks due to the mobility of DBSs. Although several works in the literature were motivated by the mobility of DBSs \cite{C_Morteza_3GPP_2019, J_Sharma_Coverage_2019, J_Enayati_mobile_2018, C_Morteza_Fundamentals_2019, J_Lyu_Cyclical_2016}, there have only been a few works that have accounted for handover in drone networks \cite{C_Park_Optimal_2015, C_Arshad_Integrating_2018}. The authors of \cite{C_Park_Optimal_2015} proposed an optimal coverage decision algorithm for seamless handover of a 3D drone network. Using the results of \cite{J_Bao_Stochastic_2015}, the trade-off between average throughput and handover rate in a multi-tier network has been studied in \cite{C_Arshad_Integrating_2018}. While these works provide useful insights, the analytical characterization of handover probability in drone networks is still an open problem, which is the main focus of this letter. More details of our contribution are provided next.

{\em Contributions.} We model the initial positions of the DBSs by a homogeneous 2D PPP and assume that they move along straight lines in random directions at a constant height. Assuming that a typical UE is served by its closest DBS, we propose two scenarios for the speed of DBSs, i.e., (i) SSM, where all DBSs move with the same speed, and (ii) DSM, where DBSs move with different speeds. We then analytically characterize the handover probability for both the SSM and the DSM. Specifically, in the SSM, we establish equivalence in the spatial distributions of the mobile DBSs as seen by a static ground UE and of the static terrestrial BSs as seen by a mobile UE. Using this equivalence, we rigorously show that the handover probability in the SSM is the same as that of a mobile UE in a single-tier cellular network studied in \cite{J_Sadr_Handoff_2015, J_Shankar_Spatio_2017}. In the DSM, we first characterize the point process of the non-serving DBSs as a function of time and then derive a lower bound on the handover probability. To the best of our knowledge, this is the first work that provides a concrete mathematical treatment of the handover probability in drone cellular networks.

\vspace{-0.2cm}
\section{System Model} \label{sec:SysMod}
We consider a network of mobile DBSs deployed at a constant height $h$ that serves the UEs on the ground. We assume that the ground is aligned with the $xy$-plane of the Cartesian coordinate system and the DBSs are located in the $z=h$ plane, which will be referred to as the DBS plane in this letter. We assume that DBSs are initially distributed as a homogeneous PPP $\Phi_{\rm D}(0)$ with density $\lambda_0$ in the DBS plane. UEs are distributed as an independent homogeneous PPP $\Phi_{\rm U}$ on the ground. We denote the origin and its projection onto the DBS plane by $\nbo = (0, 0, 0)$ and $\nbo' = (0, 0, h)$, respectively. The analysis will be performed for a \emph{typical} UE placed at $\nbo$. The distance of a DBS located at $\nbx(t) \in \Phi_{\rm D}(t)$ at time $t$ from $\nbo'$ is denoted by $u_\nbx(t) = \|\nbx(t) - \bf{o'}\|$. Moreover, we denote the location of the closest DBS to the origin and its corresponding distance to $\nbo'$ at time $t$ by ${\nbx^*}(t)$ and ${u^*}(t)$, respectively. For simplicity, we assume ${u^*} \triangleq {u^*}(0)$, and we drop $t$ for $u_\nbx(t)$ whenever the time index can be understood from the context.

We assume that each DBS moves along a straight line and in a uniformly random direction, independently of the other DBSs, in the DBS plane. In this letter, we consider two mobility scenarios for the DBSs, namely (i) SSM, where all DBSs have the same constant speed, and (ii) DSM, where DBSs have different constant speeds. Note that the SSM closely emulates the mobility model used by the 3GPP, where drones are initially placed at uniformly random locations at a constant height and then move in uniformly random directions along straight lines with the same constant speed \cite{3gpp_36777, C_Morteza_3GPP_2019}. These simple enough random straight-line mobility scenarios can be regarded as benchmarks for evaluating more sophisticated models. We assume a nearest-neighbor association policy, in which at any time $t$, the closest DBS to the typical UE is assumed to be the serving DBS and all the other DBSs are regarded as non-serving DBSs. Furthermore, the point process of the non-serving DBSs is denoted by $\Phi_{\rm D}'(t) \equiv \Phi_{\rm D}(t)\backslash {\nbx^*}(t)$.

A handover is said to occur when the serving DBS of the typical UE changes. The event of the occurrence of at least one handover until time $t$ can be mathematically expressed as
\begin{align}\label{eq:HandoverDef}
H(t)\! :=\! \left\{\exists s < t\!:\! \underset{i\in \nbbN}{\argmin} \|\nbx_i(s)\| \ne \underset{i\in \nbbN}{\argmin} \|\nbx_i(t)\|\right\},
\end{align}
where $i \in \nbbN$ is an arbitrarily assigned index to each DBS. We now define the handover probability $\nbbP[H(t)]$ at time $t$ as the probability that the first handover occurs at or before time $t$.

\section{Handover Probability} \label{sec:HP}
In this section, we derive the handover probability for both mobility scenarios. To do so, we need to first characterize the point process of the DBSs for both mobility scenarios at any time $t$. The following lemma is the direct consequence of \emph{displacement theorem} for a PPP, and thus, we state it here without a proof \cite{B_Haenggi_Stochastic_2012}.
\begin{lemma} \label{lem:NoExclusion}
Let $\Phi$ be a homogeneous PPP with density $\lambda_0$. If all the points of $\Phi$ are displaced independently of each other with identically distributed displacements, then the displaced points also form a homogeneous PPP with density $\lambda_0$.
\end{lemma}

\subsection{Handover Analysis in the SSM}
We begin our analysis with the following lemma which is the consequence of having the same speed for all DBSs in the SSM.
\begin{lemma} \label{lem:Handover1}
In the SSM, let $D_0$ be the serving DBS at time $t = t_0$ and a handover occurs at time $t = t_1$, where $t_1 > t_0$, and $D_1$ becomes the serving DBS. Then, $D_0$ cannot become the serving DBS again at any time $t > t_1$.
\end{lemma}
\begin{IEEEproof}
See Appendix \ref{app:Lemma2}.
\end{IEEEproof}
Once a handover occurs, a DBS that was acting as the serving DBS will be regarded as a non-serving DBS. Lemma \ref{lem:Handover1} states that this non-serving DBS cannot become the serving DBS again under the SSM. This fact is also in accordance with single-tier terrestrial cellular networks, where the BSs are static and a reference UE is moving with a constant speed along a straight line in a uniformly random direction. In terrestrial cellular networks, the coverage footprints are characterized by Voronoi cells when the nearest-neighbor association policy is used \cite{J_Dhillon_Modeling_2012}. Hence, handover occurs when a mobile UE crosses the boundary of a Voronoi cell. Since the Voronoi cells in single-tier terrestrial cellular networks are convex polygons, a reference UE moving along a straight line enters a Voronoi cell only once. An interesting duality between the aerial and terrestrial setups mentioned above is established in the following theorem.
\begin{theorem} \label{thm:handover1}
The handover probabilities of the following two networks are equivalent:
\begin{enumerate}
\item {\bf Terrestrial model}: A network of static BSs distributed as a homogeneous PPP $\Phi_{\rm B}$ with density $\lambda_0$. The reference UE moves along a straight line with speed $v$.
\item {\bf Aerial model}: A network of mobile DBSs initially distributed as a homogeneous PPP $\Phi_{\rm D}(0)$ with density $\lambda_0$, in which DBSs follow the SSM with speed $v$. The typical UE is static.
\end{enumerate}
\end{theorem}
\begin{IEEEproof}
Let us assume that the reference UE in the terrestrial model moves in a direction $\theta \sim U[0, 2\pi)$ w.r.t. the positive $x$-axis along a straight line and denote its trajectory by $\mathbf{x}(t)$. Then, the point process of BSs w.r.t. the reference UE can be written as $\Phi_{\rm B} - \mathbf{x}(t)$. Now, observe that the performance of the reference UE in this terrestrial network is equivalent to that of a \emph{static} typical UE in an aerial network where all the DBSs move along straight lines and in the \emph{same} direction $\pi + \theta$. Denoting this point process by $\tilde{\Phi}_{\rm D}(t)$, we have $\tilde{\Phi}_{\rm D}(t) \equiv \Phi_{\rm B} - \mathbf{x}(t)$. Since $\Phi_{\rm B}$ is a homogeneous PPP, it is translation invariant, which gives $\tilde{\Phi}_{\rm D}(t) \equiv {\rm PPP}(\lambda_0)$. Furthermore, Lemma \ref{lem:NoExclusion} states that the DBS locations in our aerial model are distributed as a homogeneous PPP with density $\lambda_0$. Hence, as seen from the UE of interest at any time $t$, the BSs and DBSs in both terrestrial and aerial models follow a homogeneous PPP with density $\lambda_0$. Consequently, the two models are equivalent in distribution at any time $t$. Note that without loss of generality, one can define the handover event ``completely" in the DBS plane, and thus, the effect of height is immaterial for the handover calculation.

Now, assume that the serving DBS in the aerial model is initially located at ${\nbx^*}(0)$ and moves to ${\nbx^*}(t)$ at time $t$. From Lemma \ref{lem:Handover1} and the definition of the handover event in \eqref{eq:HandoverDef}, we observe that a handover does not occur until time $t$ in the SSM if there is no DBS in $b(\nbo', {\nbx^*}(t))$, where $b(\nbo, r)$ is a disc of radius $r$ centered at $\nbo$. Since the probability of this event depends only on the characteristics of the point process of DBSs at time $t$ (and not on its evolution over time), we conclude that the handover probability is the same for both the terrestrial and aerial models.
\end{IEEEproof}
From the duality established in Theorem \ref{thm:handover1}, it is clear that the handover probability in the SSM is the same as that of a single-tier terrestrial cellular network. Although handover probability for single-tier terrestrial cellular networks has been derived in \cite{J_Sadr_Handoff_2015}, it is not accurate and a correction has recently been proposed as a part of a tutorial on mobility-aware performance characterization of cellular networks in \cite{J_Tabassum_Fundamentals_2019}. In what follows, we state this result and propose a slightly simpler proof.
\begin{theorem} \label{thm:handover2}
In a single-tier terrestrial cellular network, let the BSs be distributed as ${\rm PPP}(\lambda_0)$ and consider a reference UE that moves along a straight line in a uniformly random direction at a constant speed $v$. Assuming a nearest-neighbor association policy, the handover probability as seen by the reference UE at time $t$ can be written as
\begin{align} \label{eq:HandoverMain1}
\nbbP[H(t)] \!=\! 1 \!-\! \frac{1}{2\pi}\!\!\int_0^{\infty}\!\!\!\! &\int_0^{2\pi}\!\!\!\! \scalebox{0.89}{$ 2\pi\lambda_0 r \exp\bigg(\!\!\!-\!\lambda_0\Big[r^2\big(\pi \!-\! \varphi_1 \!+\! \frac{1}{2}\sin(2\varphi_1)\big)$} \nonumber\\
&\scalebox{0.9}{$+ R^2\big(\pi \!-\! \varphi_2 \!+\! \frac{1}{2}\sin(2\varphi_2)\big)\Big]\bigg)$} \,{\rm d}\theta \,{\rm d}r,
\end{align}
where $\varphi_1 \!= \cos^{-1}\left(\frac{v^2t^2 + r^2 - R^2}{2vtr}\right)$, $\varphi_2 \!= \cos^{-1}\left(\frac{v^2t^2 + R^2 - r^2}{2vtR}\right)$, and $R = \sqrt{r^2 + v^2t^2 - 2rvt\cos(\theta)}$.
\end{theorem}
\begin{IEEEproof}
See Appendix \ref{app:Theorem2}.
\end{IEEEproof}
\begin{remark} \label{rem:equivalencePH}
To the best of our understanding, this is the first work that establishes the fact that the handover probability in a drone cellular network with mobile DBSs and static UEs is equivalent to that of a terrestrial network with static BSs and mobile UEs. From Theorems \ref{thm:handover1} and \ref{thm:handover2}, we conclude that the handover probability for the SSM is as given in \eqref{eq:HandoverMain1}.
\end{remark}
\vspace{-0.5cm}
\subsection{Handover Analysis in the DSM}
In this subsection, we first characterize the point process of the non-serving DBSs in the DSM. From our construction, it is clear that $\Phi_{\rm D}'(0)$ is an inhomogeneous PPP with density
\begin{equation} \label{LambdaServiceModel1}
\lambda(u_\nbx, {u^*})=\left\{\begin{matrix}
\lambda_0 & u_\nbx > {u^*}\\ 
0 & u_\nbx \leq {u^*}
\end{matrix}.\right.
\end{equation}
Note that the nearest-neighbor association policy introduces an \emph{exclusion zone}, $\ncalX = b({\bf o'}, {u^*})$, for the non-serving DBSs. Using displacement theorem, we argue that $\Phi_{\rm D}'(t)$ remains an inhomogeneous PPP and characterize its density in the following lemma.
\begin{lemma} \label{lem:MainDensity}
In the DSM, let $V$ be a non-negative random variable representing the speed of different DBSs, with the cumulative distribution function (cdf) and probability density function (pdf) of $F_V(v)$ and $f_V(v)$, respectively. Then $\Phi_{\rm D}'(t)$ will be an inhomogeneous PPP with density
\begin{align} \label{MainLambda}
\lambda(t; \,&u_\nbx, {u^*}) = \lambda_0\bigg[1 - F_V\left(\frac{{u^*} - u_\nbx}{t}\right) \,-\nonumber\\
& \int_{\frac{|{u^*} - u_\nbx|}{t}}^{\frac{{u^*} + u_\nbx}{t}} f_V(v)\frac{1}{\pi}\cos^{-1}\left(\frac{v^2 t^2 + u_\nbx^2 - {u^*}^2}{2vtu_\nbx}\right)\,{\rm d}v\bigg].
\end{align}
\end{lemma}
\begin{IEEEproof}
See Appendix \ref{app:Lemma3}.
\end{IEEEproof}
It is clear from our setup that the locations of a DBS at two different times $t_1$ and $t_2$ are not independent of each other. Therefore, the motion of DBSs will be correlated in time, because of which the exact analysis of the handover probability in the DSM is quite hard. Hence, we provide a lower bound on the handover probability for the DSM using the marginal spatial distribution of non-serving DBSs in the following theorem.
\begin{theorem} \label{thm:handover3}
In the DSM, the handover probability can be lower-bounded as
\begin{align} \label{eq:HandoverMain2}
\nbbP[H(t)] &\geq 1 - \frac{1}{2\pi}\int_0^\infty\!\!\int_0^\infty\!\!\int_0^{2\pi} 2\pi\lambda_0 {u^*} {\rm e}^{-\lambda_0 \pi {u^*}^2} f_V(v) \,\times\nonumber\\
&\exp\left[-\int_0^R 2\pi u_\nbx \lambda(t; u_\nbx, {u^*})\, {\rm d}u_\nbx\right]\, {\rm d}\theta\, {\rm d}{u^*}\, {\rm d}v,
\end{align}
where $R = \sqrt{{u^*}^2 + v^2t^2 - 2{u^*}vt\cos(\theta)}$ and $\lambda(t; u_\nbx, {u^*})$ is given by \eqref{MainLambda}.
\end{theorem}
\begin{IEEEproof}
See Appendix \ref{app:Theorem3}.
\end{IEEEproof}
Note that the SSM is a special case of the DSM, and thus, we can also derive a lower bound on the handover probability in the SSM using Theorem \ref{thm:handover3}. However, as shown in the next corollary, the lower bound given by \eqref{eq:HandoverMain2} is exact for the SSM.
\begin{cor} \label{cor:Cor1}
The handover probability in the SSM is given by
\begin{align} \label{eq:CorHandoverMainSV}
\nbbP[&H(t)] = 1 - \frac{1}{2\pi}\int_{vt}^\infty\!\!\int_0^{2\pi} 2\pi\lambda_0 {u^*} {\rm e}^{-\lambda_0 \left[\pi {u^*}^2 + Q\right]} \, {\rm d}\theta\, {\rm d}{u^*}\nonumber\\
- \,&\frac{1}{2\pi}\int_0^{vt}\!\!\int_0^{2\pi} 2\pi\lambda_0 {u^*} {\rm e}^{-\lambda_0 \left[\pi {u^*}^2 + \pi (vt - {u^*})^2 + Q\right]} \, {\rm d}\theta\, {\rm d}{u^*},
\end{align}
where $Q = \int_{|vt - {u^*}|}^R 2\pi u_\nbx \frac{1}{\pi}\cos^{-1}\left(\frac{{u^*}^2 - u_\nbx^2 - v^2 t^2}{2u_\nbx vt}\right) \, {\rm d}u_\nbx$ and $R = \sqrt{{u^*}^2 + v^2t^2 - 2{u^*}vt\cos(\theta)}$.
\end{cor}
\begin{IEEEproof}
See Appendix \ref{app:Cor1}.
\end{IEEEproof}
Note that although the integrands of \eqref{eq:HandoverMain1} and \eqref{eq:CorHandoverMainSV} are different, the result of the integrals is the same for all $t$.

\begin{figure}[t!]
	\centering
	\includegraphics[width=0.76\columnwidth]{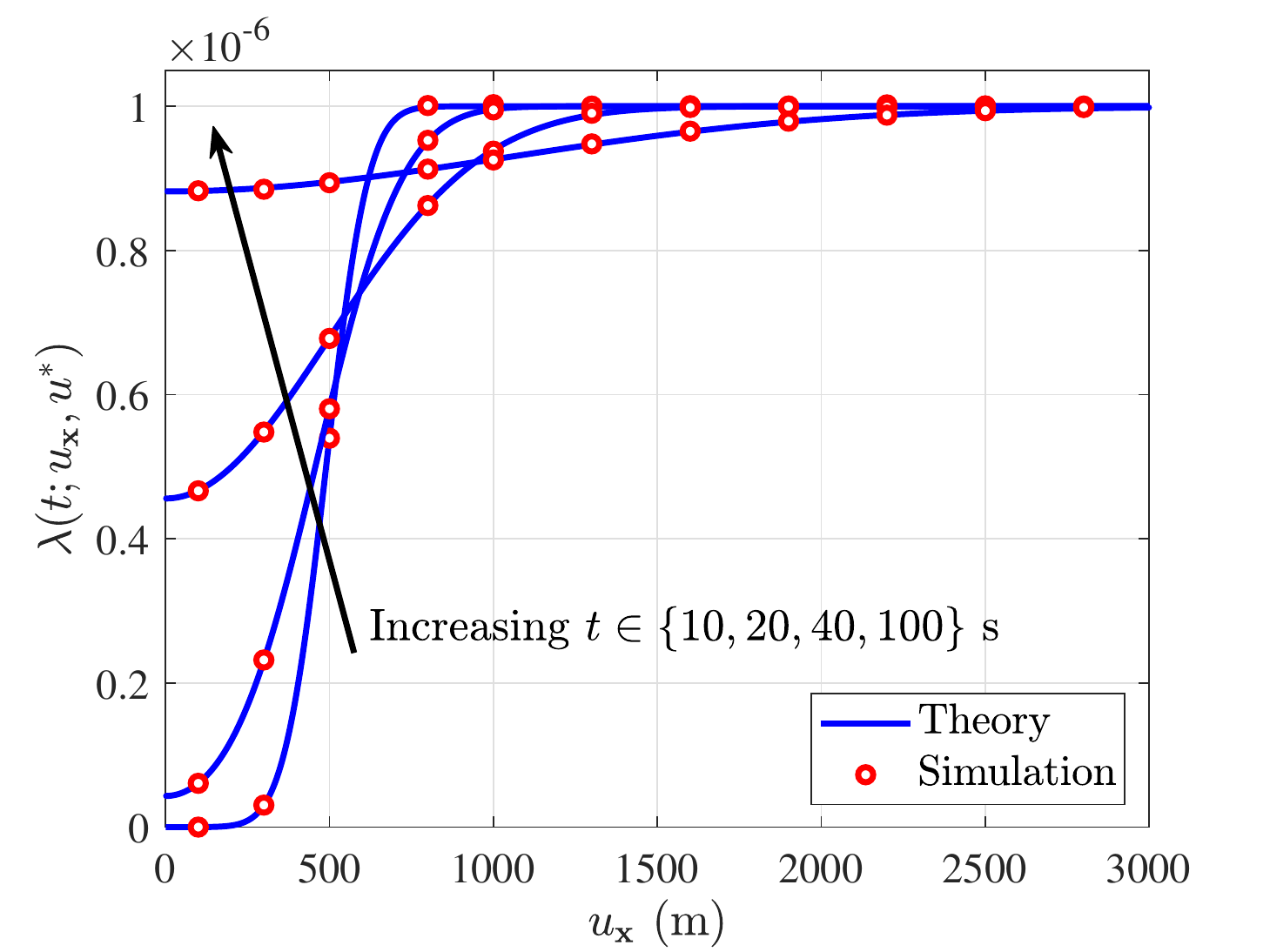}
	\vspace{-0.25cm}
	\caption{Density of non-serving DBSs for the DSM with Rayleigh distributed speed.}
	\vspace{-0.4cm}
	\label{fig:DensityPlot}
\end{figure}

\section{Simulation Results} \label{sec:Sim}
In this section, we verify the accuracy of our exact results and the proposed lower bound using Monte Carlo simulations. We assume $\lambda_0 = 1~{\rm DBS/km^2}$, $v = 45~{\rm km/h}$, and that the speed in the DSM has Rayleigh or uniform distributions with mean $v$. In Fig. \ref{fig:DensityPlot}, we plot the density of non-serving DBSs for $t \in \{10, 20, 40, 100\}~{\rm s}$ and ${u^*} = 500~{\rm m}$ in the DSM using Lemma \ref{lem:MainDensity}. Clearly, as $t \to \infty$, the point process of non-serving DBSs becomes homogeneous.

\begin{figure}[t!]
	\centering
	\includegraphics[width=0.76\columnwidth]{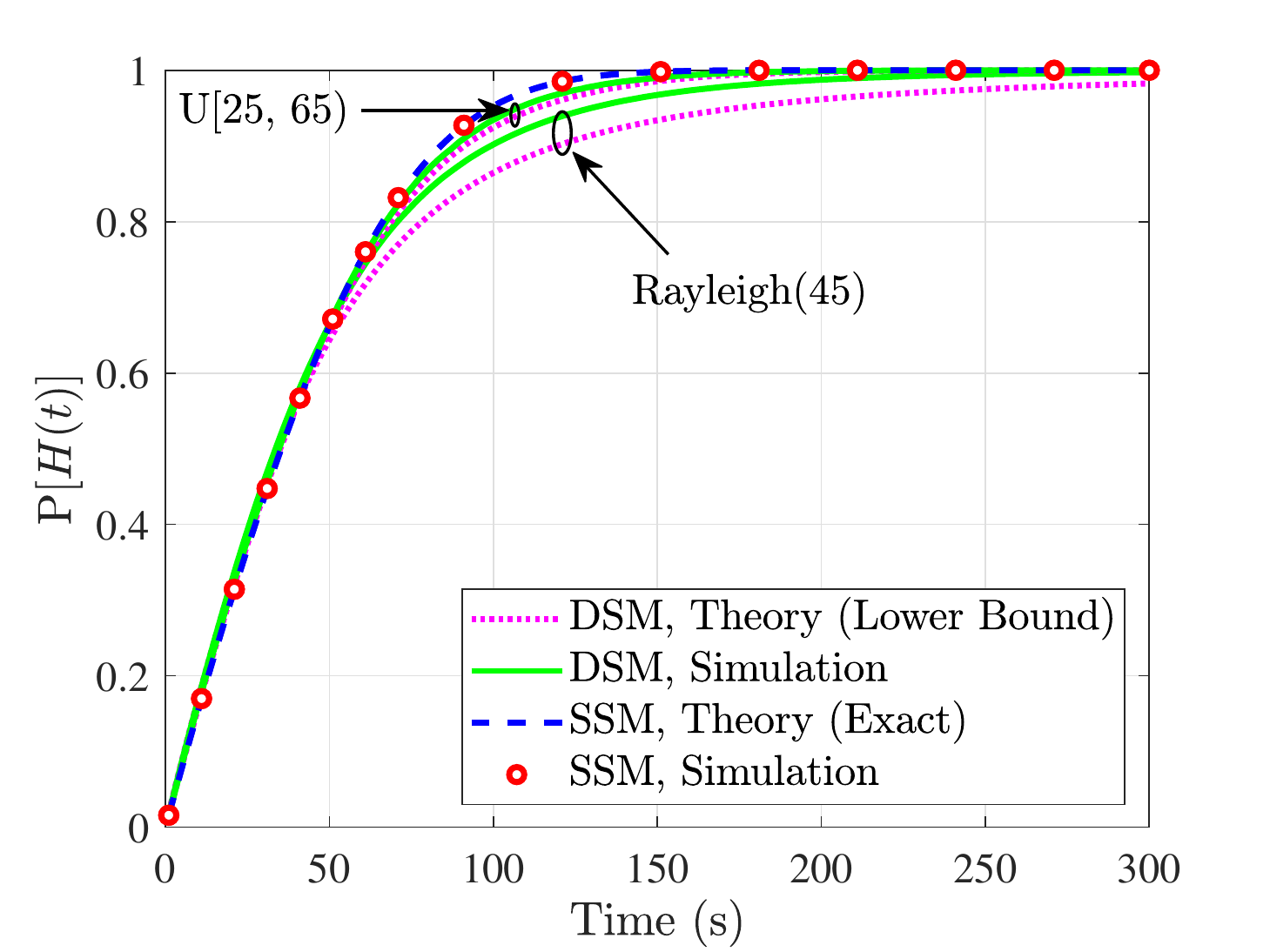}
	\vspace{-0.25cm}
	\caption{Handover probability for both mobility scenarios.}
	\vspace{-0.4cm}
	\label{fig:HandoverPlot}
\end{figure}

The handover probabilities for both the SSM and the DSM are shown in Fig. \ref{fig:HandoverPlot}. As evident from this figure, at small values of time, the handover probability is almost the same for both the mobility scenarios and the derived lower bound is tight. As the network evolves with time, we observe that the handover probability in the DSM is smaller than that of the SSM. Note that since we have defined the handover probability in Section \ref{sec:SysMod} as the probability of the occurrence of the first handover, this result does not necessarily mean that the \emph{handover rate}, defined as the average number of handovers per unit time, in the DSM will be smaller than that of the SSM.

\section{Conclusion}
In this letter, we provided a concrete mathematical treatment of the handover probability in drone cellular networks. Assuming that DBSs move along straight lines and in uniformly random directions, we considered two mobility scenarios for the DBSs: (i) SSM, where all DBSs move with the same speed, and (ii) DSM, where DBSs have different speeds. We then established a duality in terms of the handover probability between the SSM and a terrestrial network where all the BSs are static and a reference UE moves along a straight line. For the DSM, we characterized the point process of non-serving DBSs, using which we derived a lower bound for the handover probability of the network. A meaningful extension of this work could be to characterize the handover rate and other metrics directly affected by the mobility of DBSs. Another direction of research is to consider more sophisticated mobility models, e.g., where the DBSs follow a random waypoint mobility model \cite{J_Morteza_Performance_2019} or a simple cyclical mobility pattern \cite{J_Lyu_Cyclical_2016}.

\begin{figure*}[!t]
	\centering
	\subfigure[Trajectories of $D_0$ and $D_1$. Arrows denote the direction of movements of DBSs.]
	{
		\includegraphics[width=0.23\textwidth]{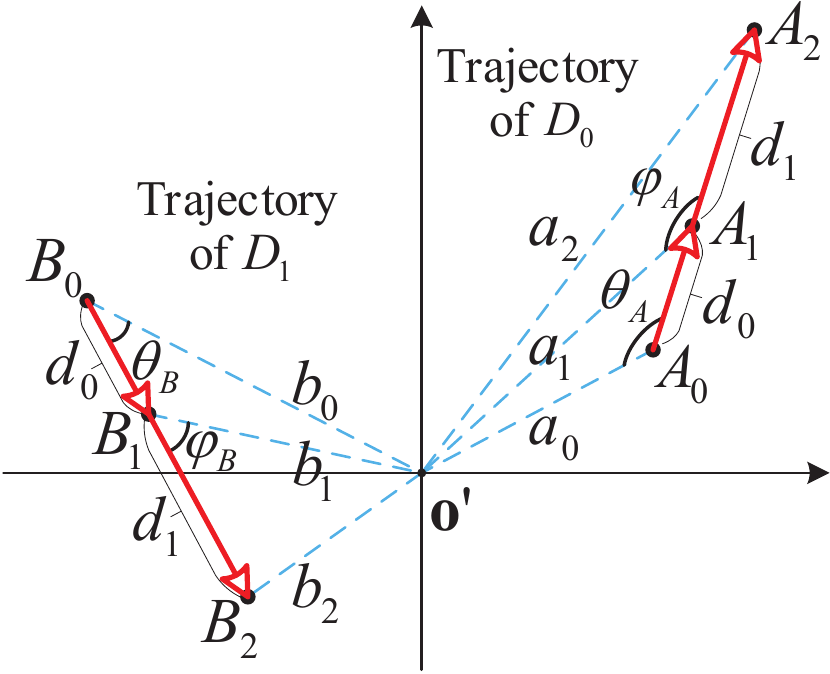}
		\label{fig:handover1}
	}\hspace{1.2cm}
	\subfigure[Movement of the reference UE. Triangles and squares denote BSs and the reference UE, respectively.]
	{
		\includegraphics[width=0.21\textwidth]{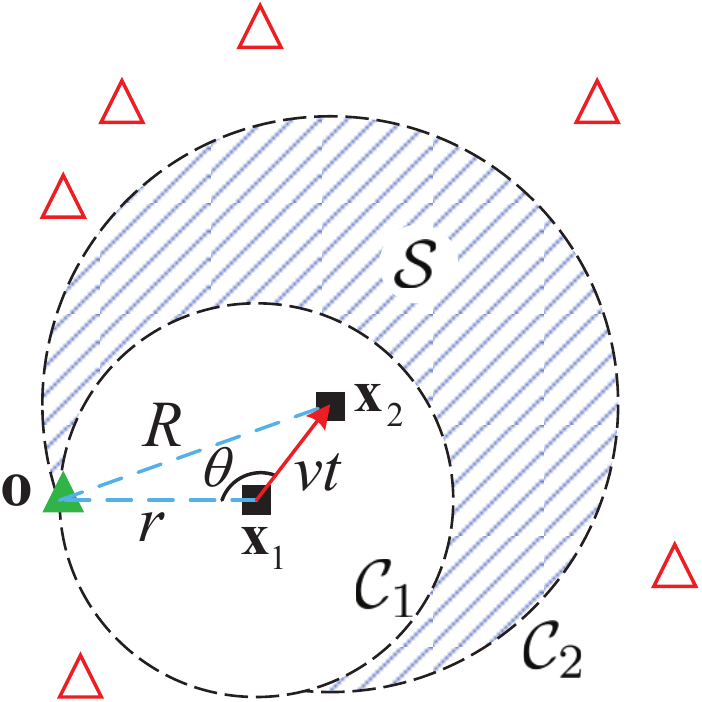}
		\label{fig:handover3}
	}\hspace{1.2cm}
	\subfigure[Movement of the DBSs. Triangles and the square denote DBSs and the typical UE, respectively.]
	{
		\includegraphics[width=0.21\textwidth]{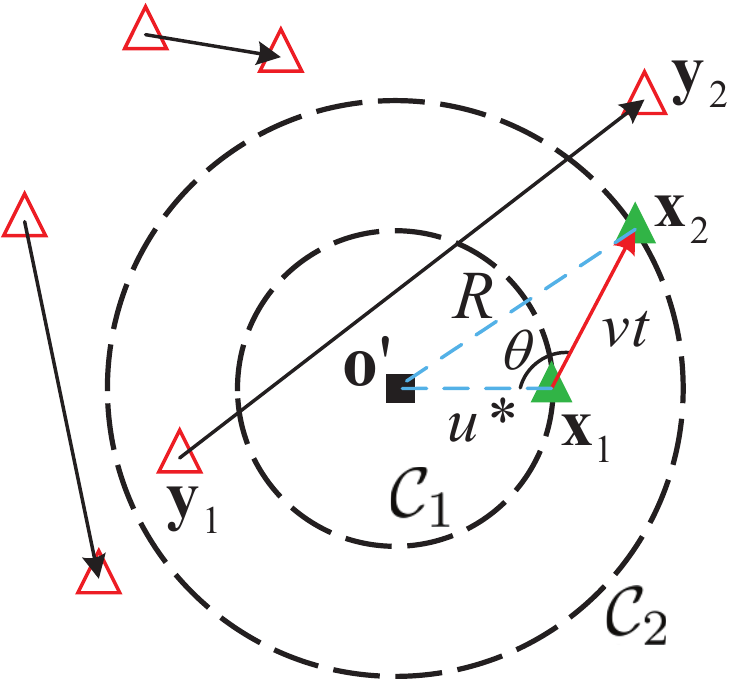}
		\label{fig:handover4}
	}
	\caption{Illustrations for the proof of (a) Lemma \ref{lem:Handover1}, (b) Theorem \ref{thm:handover2}, and (c) Theorem \ref{thm:handover3}.}
	\vspace{-0.4cm}
\end{figure*}

\appendix
\subsection{Proof of Lemma \ref{lem:Handover1}} \label{app:Lemma2}
In Fig. \ref{fig:handover1}, we represent two DBSs $D_0$ and $D_1$ and their trajectories in the time interval $[t_0, t_2]$. We denote the location of $D_0$ ($D_1$) at times $t_0$, $t_1$, and $t_2$ by $A_0$ ($B_0$), $A_1$ ($B_1$), and $A_2$ ($B_2$), respectively, and its corresponding distance from $\nbo'$ by $a_0$ ($b_0$), $a_1$ ($b_1$), and $a_2$ ($b_2$), respectively, where we assume that a handover occurs at time $t = t_1$, where $t_0 < t_1 < t_2$. Now, we need to show $a_2 > b_2$ given $a_0 < b_0$ and $a_1 > b_1$. Define $\theta_A = \angle \mathbf{o}'A_0A_2$, $\varphi_A = \angle \mathbf{o}'A_1A_2$, $\theta_B = \angle \mathbf{o}'B_0B_2$, and $\varphi_B = \angle \mathbf{o}'B_1B_2$. Without loss of generality, we assume that $\pi > \theta_A > \theta_B$. We now consider two cases:

{\bf Case 1:} $\theta_B > \frac{\pi}{2}$. We first show that $\varphi_A > \varphi_B > \frac{\pi}{2}$. From $\triangle{\mathbf{o}'B_0B_1}$, it is clear that $\varphi_B > \theta_B > \frac{\pi}{2}$. Applying the sine law in $\triangle{\mathbf{o}'A_0A_1}$ and $\triangle{\mathbf{o}'B_0B_1}$, we have
\begin{align*}
\frac{a_0}{\sin(\pi - \varphi_A)} = \frac{a_1}{\sin(\theta_A)}, \hspace{0.7cm}
\frac{b_0}{\sin(\pi - \varphi_B)} = \frac{b_1}{\sin(\theta_B)}.
\end{align*}
Since by assumption $\sin(\theta_A) < \sin(\theta_B)$ and $\frac{a_0}{a_1} < \frac{b_0}{b_1}$, we conclude that $\sin(\pi - \varphi_A) < \sin(\pi - \varphi_B)$ which gives $\varphi_A > \varphi_B > \frac{\pi}{2}$. Now writing the cosine law in $\triangle{\mathbf{o}'A_1A_2}$ and $\triangle{\mathbf{o}'B_1B_2}$, we get $a_2^2 = a_1^2 + d_1^2 - 2a_1d_1\cos(\varphi_A)$ and $b_2^2 = b_1^2 + d_1^2 - 2b_1d_1\cos(\varphi_B)$. Since $a_1 > b_1$ and $\cos(\varphi_A) < \cos(\varphi_B) < 0$, we end up with $a_2 > b_2$.

{\bf Case 2:} $\theta_B < \frac{\pi}{2}$. Writing the cosine law in $\triangle{\mathbf{o}'A_0A_1}$, $\triangle{\mathbf{o}'B_0B_1}$, $\triangle{\mathbf{o}'A_0A_2}$, and $\triangle{\mathbf{o}'B_0B_2}$, we have
\begin{align*}
a_1^2 &= \scalebox{0.96}{$ a_0^2 + d_0^2 - 2a_0d_0\cos(\theta_A), \hspace{0.22cm}b_1^2 = b_0^2 + d_0^2 - 2b_0d_0\cos(\theta_B),$}\\
a_2^2 &= a_0^2 + (d_0+d_1)^2 - 2a_0(d_0+d_1)\cos(\theta_A),\\
b_2^2 &= b_0^2 + (d_0+d_1)^2 - 2b_0(d_0+d_1)\cos(\theta_B).
\end{align*}
Now, to show the inequality $a_2 > b_2$, we can write
\begin{align*}
a_2^2 \!>\! b_2^2 \!&\Longleftrightarrow\! \scalebox{0.94}{$ a_0^2 \!-\! 2a_0(d_0\!+\!d_1)\cos(\theta_A) \!>\! b_0^2 \!-\! 2b_0(d_0\!+\!d_1)\cos(\theta_B)$}\\
&\hspace{-0.8cm}\Longleftrightarrow \,a_1^2 - 2a_0d_1\cos(\theta_A) > b_1^2 - 2b_0d_1\cos(\theta_B)\\
&\hspace{-0.8cm}\Longleftrightarrow \,(a_1^2 - b_1^2) + 2d_1(b_0 \cos(\theta_B) - a_0 \cos(\theta_A)) > 0.
\end{align*}
The last inequality is valid since $\cos(\theta_B) > \max\{0, \cos(\theta_A)\}$ and $b_0 > a_0$, which gives $b_0 \cos(\theta_B) > a_0 \cos(\theta_A)$.
\hfill 
\IEEEQED

\subsection{Proof of Theorem \ref{thm:handover2}} \label{app:Theorem2}
Consider the set of BSs and the reference UE in Fig. \ref{fig:handover3}, where the serving BS is located at $\mathbf{o}$ and the reference UE moves a distance of $vt$ in a uniformly random direction $\theta$ from $\mathbf{x}_1$ to $\mathbf{x}_2$. The distance of the reference UE from $\mathbf{o}$ before and after its movement is $r$ and $R$, respectively. Defining $\mathcal{C}_1 = b(\mathbf{x}_1, r)$ and $\mathcal{C}_2 = b(\mathbf{x}_2, R)$ as two open balls, handover will not occur if there is no BS in $\mathcal{C}_2$. Since $\mathcal{C}_1$ is empty by definition, handover will not occur if there is no BS in $\mathcal{C}_2 \backslash \mathcal{C}_1$ (shaded region in Fig. \ref{fig:handover3}). Hence, the handover probability can be written by conditioning on $r$ and $\theta$ as
\begin{align} \label{eq:PH_Condition}
\nbbP[H(t)|r, \theta]\! &= \scalebox{0.95}{$\! 1 \!-\! \nbbP\left[N(\mathcal{C}_2 \backslash \mathcal{C}_1) \!=\! 0\right] \!=\! 1 \!-\! \nbbP\left[N(\mathcal{C}_2 \backslash (\mathcal{C}_1 \!\cap\! \mathcal{C}_2)) \!=\! 0\right]$}\nonumber\\
&\overset{(*)}{=} 1 \!- {\rm e}^{-\lambda_0|\mathcal{C}_2 \backslash (\mathcal{C}_1 \cap \mathcal{C}_2)|} = 1 \!- {\rm e}^{-\lambda_0(\pi R^2 - \ncalS)},
\end{align}
where $N(B)$ denotes the number of points in set $B$, step $(*)$ follows from the null probability of PPP($\lambda_0$), and $\ncalS$ is the intersection area between $\mathcal{C}_1$ and $\mathcal{C}_2$, which can be written from plane geometry as
\begin{align} \label{eq:IntersectArea}
\ncalS = r^2 \left( \varphi_1 - \frac{1}{2}\sin(2\varphi_1) \right) + R^2 \left( \varphi_2 - \frac{1}{2}\sin(2\varphi_2) \right),
\end{align}
where $R$, $\varphi_1$, and $\varphi_2$ are as given in the theorem statement. Substituting \eqref{eq:IntersectArea} into \eqref{eq:PH_Condition} and deconditioning on $r$ and $\theta$ gives the final result for the handover probability as in \eqref{eq:HandoverMain1}.
\hfill 
\IEEEQED

\subsection{Proof of Lemma \ref{lem:MainDensity}} \label{app:Lemma3}
The non-serving DBSs are initially distributed as an inhomogeneous PPP with density given in \eqref{LambdaServiceModel1}. Since the displacements are independent of each other in the DSM, the resulting network at time $t$ will also be an inhomogeneous PPP because of displacement theorem \cite{B_Haenggi_Stochastic_2012}. Lemma \ref{lem:NoExclusion} asserts that without the exclusion zone $\ncalX$, DBSs will be distributed as ${\rm PPP}(\lambda_0)$ at any time $t$. However, in the presence of $\ncalX$, we can partition the set of non-serving DBSs into two sets: (i) non-serving DBSs initially inside $\ncalX$, and (ii) non-serving DBSs initially outside $\ncalX$. We denote the density due to the former and latter by $\lambda_1(t; u_\nbx, {u^*})$ and $\lambda(t; u_\nbx, {u^*})$, respectively. Note that $\lambda(t; u_\nbx, {u^*})$ is the density of the network of non-serving DBSs and we have $\lambda(t; u_\nbx, {u^*}) = \lambda_0 - \lambda_1(t; u_\nbx, {u^*})$. Using the same treatment as in the proof of Lemma 2 in \cite{J_Morteza_Performance_2019}, we get $\lambda_1(t; u_\nbx, {u^*}) =$
\begin{align} \label{Lambda1First}
\frac{\lambda_0}{\pi}{\int_0^{\infty}}\!\!\!\int_{\ncalR_1}\frac{2r f_V(v)}{\sqrt{(u_\nbx^2 - (vt-r)^2)((vt+r)^2 - u_\nbx^2)}}\,{\rm d}r\,{\rm d}v,
\end{align}
where $\ncalR_1 = \left\{|vt-u_\nbx| \leq r \leq vt+u_\nbx\right\} \bigcap \left\{0 \leq r \leq {u^*}\right\}$. Note that $\ncalR_1$ can be simplified further by considering the relations between ${u^*}$, $|vt-u_\nbx|$, and $vt+u_\nbx$. We skip further details for brevity. We finally get $\lambda_1(t; u_\nbx, {u^*}) = $
\begin{align} \label{MainLambda1}
\scalebox{0.87}{$\lambda_0\bigg[F_V\!\left(\!\frac{{u^*} \!- u_\nbx}{t}\!\right) $} \!+\!\! \int_{\frac{|{u^*} \!- u_\nbx|}{t}}^{\frac{{u^*} \!+ u_\nbx}{t}} \scalebox{0.90}{$ \!f_V(v)\frac{1}{\pi}\cos^{-1}\!\left(\frac{v^2 t^2 + u_\nbx^2 - {u^*}^2}{2vtu_\nbx}\right){\rm d}v \bigg]$},
\end{align}
which gives the density of non-serving DBSs as in \eqref{MainLambda}.
\hfill 
\IEEEQED

\vspace{-0.3cm}
\subsection{Proof of Theorem \ref{thm:handover3}} \label{app:Theorem3}
According to Fig. \ref{fig:handover4}, let $\nbx_1$ be the initial location of the serving DBS with distance ${u^*}$ from $\nbo'$. Assume that the serving DBS moves to a new location $\nbx_2$ by time $t$ with speed $v$ in direction $\theta$. Let $R = \|\nbo'\nbx_2\|$, $\mathcal{C}_1 = b(\nbo', {u^*})$, and $\mathcal{C}_2 = b(\nbo', R)$. Defining $G$ as the event that there is no DBS in $\ncalC_2$ and $\bar{H}(t)$ as the event that handover has not been occurred until time $t$, we have $\bar{H}(t) \subset G$. This is due to different speeds of DBSs and the probable event that a DBS can enter and exit $\ncalC_2$ before time $t$ (see the movement of a non-serving DBS from $\nby_1$ to $\nby_2$ in Fig. \ref{fig:handover4}). Hence, $\nbbP[\bar{H}(t)] \leq \nbbP[G]$, which gives
\begin{align*}
\nbbP[H(t)] &= 1 - \nbbP[\bar{H}(t)] \geq 1 - \nbbP[G] = 1 - \nbbP[N(\ncalC_2) = 0] \\
&= 1 - \exp\left[-\int_0^R 2\pi u_\nbx \lambda(t; u_\nbx, {u^*})\, {\rm d}u_\nbx\right].
\end{align*}
Deconditioning w.r.t. $\theta$, ${u^*}$, and $v$, we end up with \eqref{eq:HandoverMain2}.
\hfill 
\IEEEQED

\subsection{Proof of Corollary \ref{cor:Cor1}} \label{app:Cor1}
In the SSM, since we assume all DBSs have the same speed $v$, we have $f_V(v') = \delta(v' - v)$. Substituting this equation into \eqref{eq:HandoverMain2}, we end up with \eqref{eq:CorHandoverMainSV} with some mathematical manipulations. Now based on Lemma \ref{lem:Handover1}, events $G$ and $\bar{H}(t)$ (defined in Appendix \ref{app:Theorem3}) will become equivalent, i.e., the occurrence of a handover before time $t$ necessitates the existence of a point in $\ncalC_2$. Hence, the derived lower bound in \eqref{eq:HandoverMain2} becomes exact in \eqref{eq:CorHandoverMainSV} and the proof is complete.
\hfill 
\IEEEQED
\vspace{-0.1cm}

{ \setstretch{0.97}
\bibliographystyle{IEEEtran}
\bibliography{J2}
}
\end{document}